\documentclass{article}
\usepackage{spconf}
\usepackage{graphicx}
\usepackage[normalem]{ulem}
\usepackage{bm} 
\usepackage{multirow}
\usepackage{colortbl}

\usepackage{cite}
\usepackage{amsmath,amssymb,amsfonts} %
\usepackage{algorithmic}
\usepackage{textcomp}
\usepackage{xcolor}

\title{LGSQE: Lightweight Generated Sample Quality Evaluation}

\name{Ganning~Zhao \textsuperscript{1}, 
Vasileios Magoulianitis \textsuperscript{1}, 
Suya You \textsuperscript{2},
C.-C.~Jay~Kuo \textsuperscript{1}
}

\address{University of Southern California, Los Angeles, California, USA$^1$\\
DEVCOM Army Research Laboratory, Los Angeles, CA 90094, USA$^2$}

\begin{document}

\ninept
\maketitle

\begin{abstract}

Despite prolific work on evaluating generative models, little research
has been done on the quality evaluation of an individual generated
sample. To address this problem, a lightweight generated sample quality
evaluation (LGSQE) method is proposed in this work.  In the training
stage of LGSQE, a binary classifier is trained on real and synthetic
samples, where real and synthetic data are labeled by 0 and 1,
respectively. In the inference stage, the classifier assigns soft labels
(ranging from 0 to 1) to each generated sample. The value of soft label
indicates the quality level; namely, the quality is better if its soft
label is closer to 0. LGSQE can serve as a post-processing module for
quality control. Furthermore, LGSQE can be used to evaluate the
performance of generative models, such as accuracy, AUC, precision and
recall, by aggregating sample-level quality. Experiments are conducted
on CIFAR-10 and MNIST to demonstrate that LGSQE can preserve the same
performance rank order as that predicted by the Fréchet Inception
Distance (FID) but with significantly lower complexity. 
\end{abstract}
\section{Introduction}\label{sec:introduction}

Evaluation of image generative models has become an active research
topic due to the rapid advances in adversarial and non-adversarial
generative models.  Advanced image generative models find applications
in image generation, image inpainting, image-to-image translation, etc.
Due to the popularity of generative models, it is essential to have an
automatic means to measure the quality of generated samples in an
objective way without involving the human subjective test. 

Quite a few quantitative metrics have been proposed in recent years
\cite{borji2019pros, borji2022pros}. Examples include the Inception
Score (IS) \cite{salimans2016improved}, the Fréchet Inception Distance
(FID) \cite{heusel2017gans}, the Classifier two-sample test
\cite{lopez2016revisiting} and the Precision and Recall (P\&R)
\cite{sajjadi2018assessing}, etc. Each metric has its own strengths and
weaknesses. Although being popular, the aforementioned evaluation
methods share two common issues.  First, they measure the effectiveness
of a generative model based on the statistics of its whole generated
samples. They cannot be applied to a single generated sample. In
certain applications, it is desired to assess the quality of individual
samples on the fly in the absence of ensemble distributions. Second, an
important aspect for an evaluation method is its computational
complexity. Some methods rely on deep features obtained from late layers
of deep neural networks (DNNs). As a result, their computational
complexity and memory cost are high. Also, certain evaluation methods
are biased towards the ImageNet that is commonly adopted in pre-trained
networks. Although efforts have been made to develop better quality
evaluation methods, e.g.  \cite{alaa2022faithful, im2018quantitatively,
naeem2020reliable}, these two fundamental problems still exist. 

A lightweight generated sample quality evaluation (LGSQE) method is
proposed to address them in this work. LGSQE trains a binary classifier
to differentiate real and synthetic samples generated by a generative
model. In the training stage, real and generated samples are assigned
with labels ``zero" and ``one", respectively.  In the testing stage, a
soft label is obtained, which serves as its quality index. The sample
quality is good (or bad) if its soft label is farther away from (or
close to) one. The LGSQE pipeline consists of three steps: 1) design a
simple yet effective representation for real/synthetic images from a
source dataset, 2) determine discriminant features, and 3) conduct
binary classification.  

As a byproduct, LGSQE can provide quality metrics for a generative model
by aggregating quality indices of a large number of generated samples.
Intuitively, a poorly-performing generative model tends to yield more bad
samples. The distribution of generated samples from a poorly-performing
generative model is quite different from that of real samples in the
sample representation space. The accuracy of the binary classifier is
higher since their distributions are more separable. In contrast, if the
classification performance is close to chance level (i.e., half-half),
it indicates a high-performing model that generates high-quality samples
that are close to the real ones in the representation space.  LGSQE is a
dataset-specific method. The dataset chosen as the generation target
(e.g., CIFAR-10, etc.) is used to train LGSQE from scratch, as implemented
in Step 1. It is worth noting existing quality metrics for generative
models are all not dataset-specific. 

The rest of this paper is organized as follows. Related work is briefly
reviewed in Sec. \ref{sec:review}. The LGSQE method is presented in Sec.
\ref{sec:method}. Experimental results are shown in Sec.
\ref{sec:experiments}. Concluding remarks and future research directions
are given in Sec. \ref{sec:conclusion}. 

\section{Related Work}\label{sec:review}

Quite a few metrics for generative model evaluation have been proposed.
They are reviewed in Sec.  \ref{subsec:metrics}. Furthermore, we conduct
a brief survey on recent development in green learning in Sec.
\ref{subsec:green}, as it pertains to the representation learning and
feature selection of the proposed LGSQE method. 

\subsection{Evaluation Metrics for Generative Models}\label{subsec:metrics}

The Inception Score (IS) \cite{salimans2016improved} is one of the early
developed metrics. It uses the Inception-Net pre-trained on ImageNet to
calculate the KL-divergence between the conditional and marginal
distributions. It has some limitations. First, it is susceptible to
overfit \cite{yang2017lr}. Second, it fails to account for the mode
collapse problem with generative models and its bias towards
ImageNet may give an image quality assessment in an object-wise manner
(rather than realistically-wise one). Third, it is sensitive to the image
resolution and not being a proper distance metric. 

The Fréchet Inception Distance (FID) \cite{heusel2017gans} is meant to
improve deficiencies of IS. Inception-V3 is used to map samples onto
its embedding space, where real and synthetic samples are modeled by
joint Gaussian distributions. FID improves over IS by accounting for
intra-class mode dropping and in turn the diversity of generated samples
between models. Yet, the log-likelihood distributions between real and
synthesized samples are not easy to be captured in the high dimensional
feature space \cite{theis2015note}. FID is further enhanced in
\cite{liu2018improved} by introducing the Class-Aware Frechet Distance
(CAFD). 

The precision-recall metric with a reference data manifold was
introduced in \cite{lucic2018gans}. It attempts to take both fidelity
and diversity into account. Yet, it has a bottleneck in real
applications; namely, it is impractical to be deployed since the
reference manifold is not available in most settings. Other reported
limitations include failure to realize the match between identical
distributions and robustness to outliers \cite{naeem2020reliable}. 

Another line of research adopts classifier-based evaluation
\cite{im2018quantitatively} by training a classifier on real and
synthetic samples. The classifier plays the role of a discriminator, and
its error rate is used for performance assessment. For instance, the
two-sample test \cite{lopez2016revisiting} adopts the k-nearest neighbor
(KNN) classifier trained on deep-layer embeddings from a third-party DNN
classifier. 

Aforementioned evaluation metrics target to evaluate generative models.
Little research has been conducted on the quality evaluation of
individual generated samples. Sample-based evaluation can be used to
select faithful samples and reject those of lower fidelity for quality
control. LGSQE offers a two-fold solution by enabling both sample-based
and ensemble-based evaluations. 

\subsection{Green Learning}\label{subsec:green}

Green learning (GL) \cite{kuo2022green} aims at the design of a
lightweight learning system that has a small model size, fast training
time, and low inference complexity. It consists of unsupervised
representation learning, supervised feature learning and supervised
classification learning three modules. All of them can be done
efficiently. GL was initiated by Kuo with an effort to understand DL in
\cite{kuo2016understanding, kuo2017cnn}. Afterwards, the Saab transform
\cite{kuo2019interpretable} was proposed to find image representations
without backpropagation. The family of PixelHop methods was developed in
\cite{chen2020pixelhop, chen2020pixelhop++, yang2022design}. Apart from
representation learning, a powerful feature selection tool, called the
discriminant feature test (DFT), was proposed in
\cite{yang2022supervised}. DFT builds a bridge between representations
learned without labels and labels. LGSQE is a quality evaluation metric
based on the green learning principle. 

\section{Proposed LGSQE Method}\label{sec:method}

The LGSQE method consists of three cascaded modules as elaborated below. 

\noindent
{\em Module 1: Representation Learning} \label{subsec:repres_Saab}

In this module, effective local and global representations of images are
learned. The module may contain the processing in several stages, where
each stage is called a hop. One hop pipeline is adopted due to its high performance and low complexity. The input is 
images of size $N \times N \times C$. We consider overlapping blocks of size $F
\times F \times C$ with stride equal to $S$, where $F \times F$ is the
spatial size and $C$ is the channel number. The Saab transform
\cite{kuo2019interpretable} is applied to these blocks to learn
effective representations for downstream classification. 

The Saab transform applies the constant element vector of unit length to
an input block to get its DC coefficient. Then, it subtracts the DC
component, applies the principal component analysis (PCA) to the
residual, and derive the AC kernels as frequency-selective filters. The
AC kernels of larger eigenvalues correspond to lower frequency
components. This operation yields filter responses in the form of 3D
tensors. The 3D tensors are 2D spatial dimensions $N_1 \times N_1$,
where $N_1=(N-F)/S+1$, and 1D spectral dimension $K=F\times F\times C$.
The high frequency components with very small energy are discarded for
dimension reduction, so the actual number $K_1$ of spectral channels is
less than $K$. It is a user-determined parameter. 

The absolute max-pooling, as introduced in \cite{yang2022design}, is applied to each channel. It results in a 3D tensor $1/2 N_1 \times 1/2 N_1 \times K_1$ which is the spatial representation and is also used as the input to the next step. The channel-wise Saab transform (c/w Saab) with $F = N_1 $, as proposed in \cite{chen2020pixelhop++}, is applied to further reduce dimensions and allow larger receptive fields so that representations of longer distance correlation can be extracted effectively. It conducts the Saab transform on each channel separately. This step generates the spectral representation.
All spatial (local) and spectral (global) representations
are concatenated to build a rich representation set for discriminant
feature selection in Module 2. 

\begin{figure}[ht]
\centering 
\includegraphics[width=0.3\textwidth]{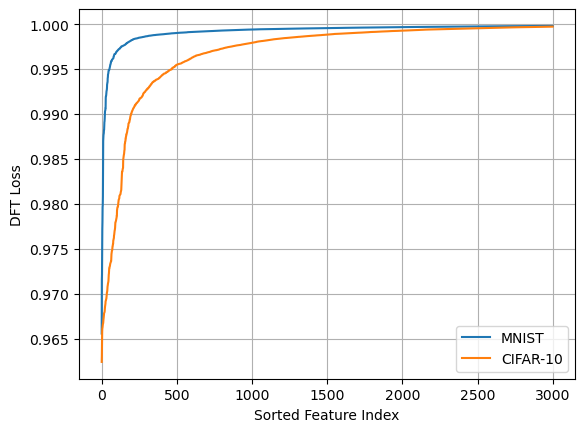}
\caption{Illustration of the DFT curves for MNIST and CIFAR-10, where
the y-axis indicates the loss function and the x-axis denotes the sorted
representation index.} \label{fig:dft_curves}
\end{figure}

\begin{figure*}[ht]
\centering 
\includegraphics[width=0.33\linewidth]{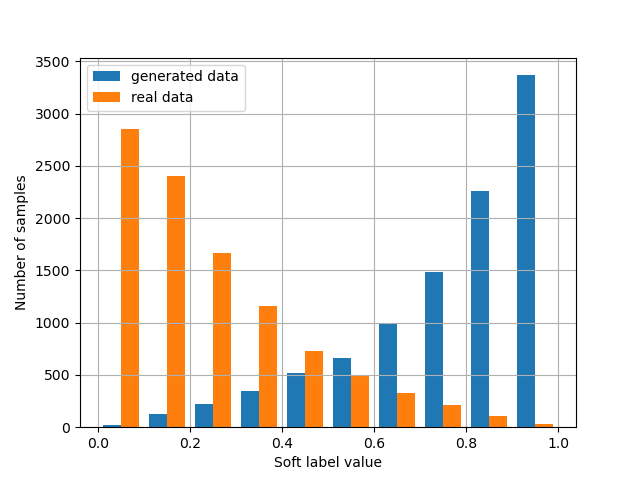}
\includegraphics[width=0.33\linewidth]{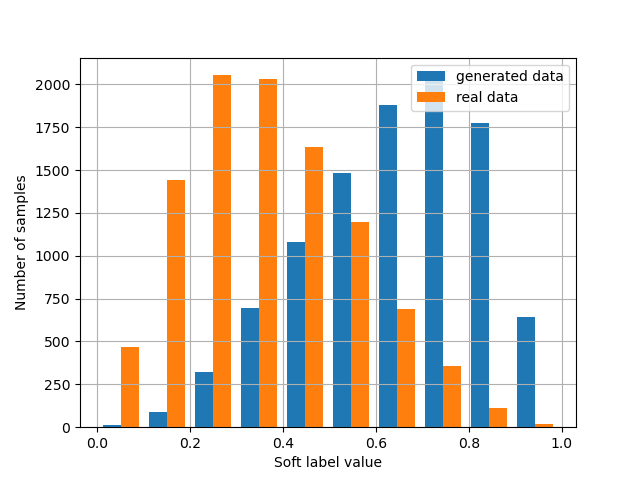} 
\includegraphics[width=0.33\linewidth]{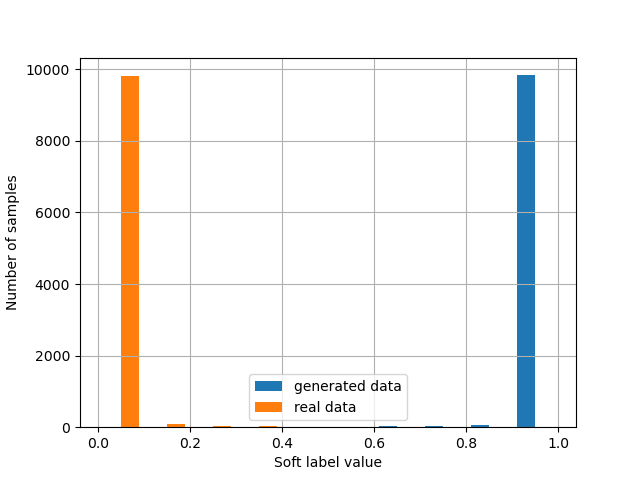}\\
(a) \hspace {5.3cm} (b) \hspace {5.3cm} (c)
\caption{The soft score histograms with generator models ``Diffusion
StyleGAN2" and ``Styleformer" for CIFAR-10 are shown in (a) and (b),
respectively, and the soft score histogram with generator model
``WGAN-GP" for MNIST is shown in (c).} \label{fig:cifar_hist}
\end{figure*}

\begin{figure}[ht]
\centering 
\includegraphics[width=0.75\linewidth]{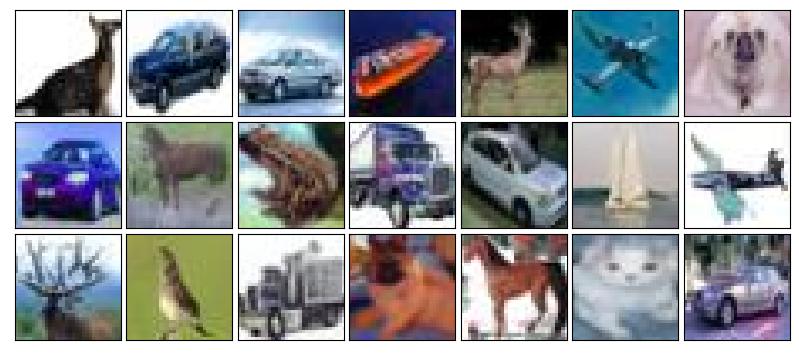}
\\ (a) \\
\includegraphics[width=0.75\linewidth]{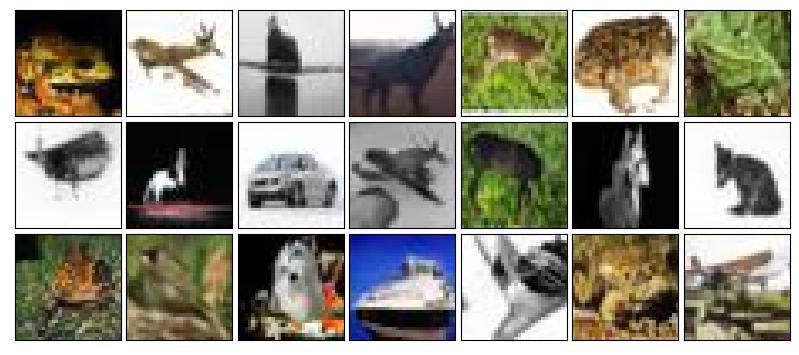}
\\ (b) \\
\caption{Visualization of generated CIFAR-10 samples where samples of
quality index close to 0 and 1 are shown in (a) and (b),
respectively.} \label{fig:examples_cifar}
\end{figure}

\begin{figure}[ht]
\centering 
\includegraphics[width=0.75\linewidth]{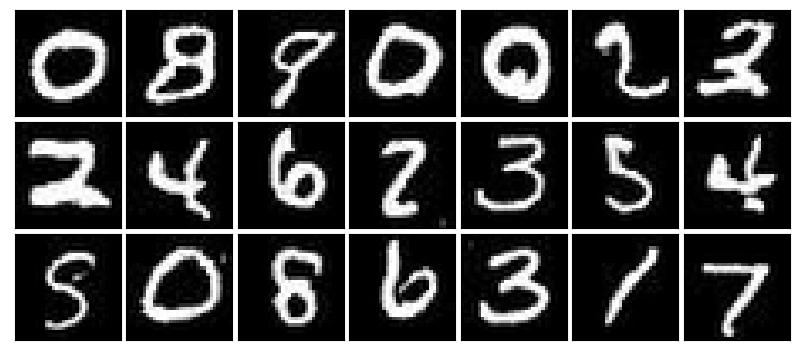}
\\ (a) \\
\includegraphics[width=0.75\linewidth]{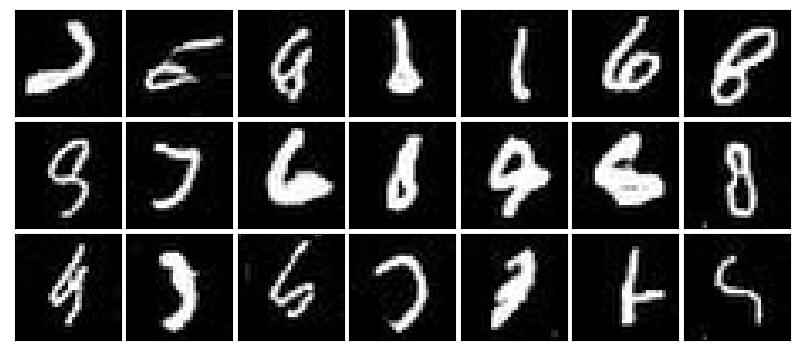}
\\ (b)\\
\caption{Visualization of generated MNIST samples where samples of
quality index close to 0 and 1 are shown in (a) and (b),
respectively.} \label{fig:examples_mnist}
\end{figure}

\noindent
{\em Module 2: Discriminant Feature Test (DFT)}

The number of representations obtained from Module 1 is large. We need a
mechanism to choose powerful ones against a particular task. This is
achieved by the discriminant feature test (DFT)
\cite{yang2022supervised}. DFT analyzes the discriminant power of each
representation based on the following idea. For the $i^{th}$
representation, it computes its value range, $[f^i_{min}, f^i_{max}]$,
and partition the range into two non-overlapping subsets, denoted by
$S^i_L$ and $S^i_R$, with a set of uniformly-spaced  partition points. The
DFT loss is defined as the smallest weighted entropy of $S^i_L$ and $S^i_R$
in left and right partitions at the optimal partition point. Mathematically,
it can be written as
\begin{equation}\label{eq:dataset}
L_{DFT} = \mathop{\min}_{t\in T} H_t^i = - \mathop{\min}_{t\in T}
\sum^C_{c=1} [p^i_{L,c}\log(p^i_{L,c}) + p^i_{R,c}\log(p^i_{R,c})]
\end{equation}
Where $T$ denotes the set of uniformly-spaced partition points, $C$ is
the class number and $p^i_{L,c}$ and $p^i_{R,c}$ denote the probability
of the $i^{th}$ representation dimension being in class $c$ of the left or
right interval, respectively. 

The DFT loss can be computed for all representations in parallel since
they are independent. The lower the DFT loss, the more discriminant of
the representation dimension. The representations are sorted by their
DFT loss in ascending order to yield the DFT loss curve as shown in
Fig. \ref{fig:dft_curves}. We can use the elbow point to select a
subset of representations with lower DFT loss. They define a set of
discriminant features to be fed into a binary classifier in Module 3. 

\noindent
{\em Module 3: Binary Classification for Evaluation}

We partition the real/generated data into training and testing sets.  A
binary classifier is trained on the union of real and generated training
samples. They are labeled with ``0" and ``1", respectively. The
classifier assigns a soft score, $0\leq d \leq 1$, to each testing
sample as the sample quality index. The hard decision depends on
threshold $t$, where $0 \leq t \leq 1$. The sample is labeled as ``real"
if $0 \leq d < t$, and ``generated" if $t \leq d \leq 1$. Four commonly
used performance metrics of a binary classifier are reviewed below. They
are accuracy (Acc.), precision (Pre.), recall (Rec.) and area under the
curve (AUC). Accuracy is the ratio of the ``correct decision number" over
the ``total decision number". Precision and Recall is defined as
\begin{equation}
\mbox{Pre} = \frac{TP}{TP+FP}, \quad \mbox{Rec} = \frac{TP}{TP+FN},
\end{equation}
where, $TP$, $FP$, $TN$ and $FN$ indicate true positive, false positive,
true negative, and false negative, respectively. AUC is computed by the
precision-recall curve by varying threshold $t$ from zero to one. 

\begin{figure}[ht]
\centering 
\includegraphics[width=0.3\textwidth]{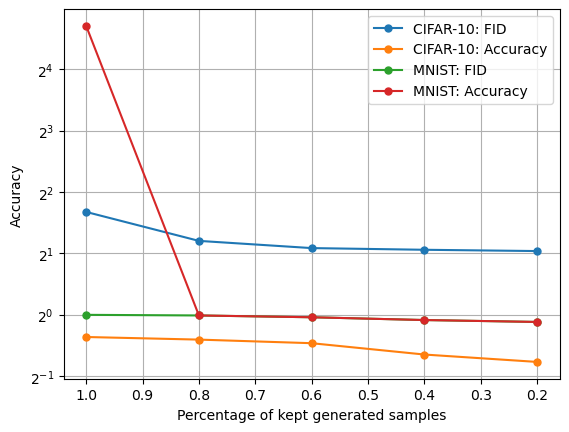}
\caption{Post-processing result.}\label{fig:discriminator}
\end{figure}

\begin{figure}[ht]
\centering 
\includegraphics[width=0.3\textwidth]{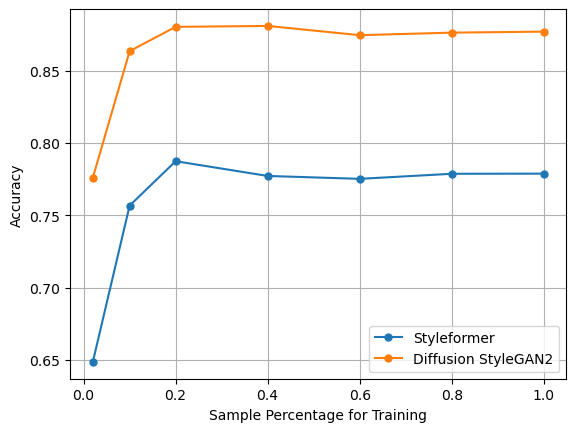}
\caption{The correct classification rates of LGSQE for samples 
generated by two models as a function of the number of real 
samples (expressed in terms of percentages of the total training
samples).}\label{fig:diff_triannum}
\end{figure}

\section{Experiments} \label{sec:experiments}

To show the effectiveness of the LGSQE method, we conduct experiments on
the CIFAR-10 dataset, which is one of the most popular datasets in the
image generation field. Multiple generative models such as DCGAN
\cite{radford2015unsupervised}, StyleFormer \cite{park2022styleformer},
StyleGAN2-ADA \cite{karras2020training}, Diffusion-StyleGAN2
\cite{wang2022diffusion}, and StyleGANXL \cite{sauer2022stylegan} have
been developed for this dataset. We also employ the MNIST dataset
\cite{lecun1998gradient} in the experiment with three generative models;
namely, GAN \cite{goodfellow2014generative}, WGAN
\cite{arjovsky2017wasserstein} and WGAN-GP \cite{gulrajani2017improved}.
The hyperparameters in Module 1 are $F=3$, $S=1$ and $C=3$ for CIFAR-10
and $F=5$, $S=2$ and $C=1$ for MNIST. After discarding Saab coefficients
of extremely low energy, 3,500 representations remain and DFT is
conducted to get 800 most discriminant features for CIFAR-10.
Similarly, 3,000 representations remain and 400 most discriminant
features are kept for MNIST in Module 2.  Finally, the XGBoost (extreme
gradient boosting) classifier \cite{chen2016xgboost} is adopted in
Module 3 as it has high performance with a small model size. 

\textbf{Evaluation of generated samples.} A soft score (the probability
of a sample belonging to class ``one") is assigned to each generated
sample as a quality index by LGSQE. If a generated sample has a score
close to one, it is likely to be a generated one.  In contrast, a soft
score closer to zero indicates a higher likelihood of being a real one.  We
show histograms of soft scores for real and generated samples computed
by LGSQE for CIFAR-10, where the generator models in Fig.
\ref{fig:cifar_hist} (a) and (b) are Diffusion StyleGAN2 and
Styleformer, respectively. By comparing these two histograms, we claim
that Styleformer is a better generator model than Diffusion StyleGAN2,
since its generated samples and real ones are clustered in the middle
region of soft score 0.5. They are more difficult to distinguish.  The
soft score histogram of WGAN-GP for MNIST is shown in Fig.
\ref{fig:cifar_hist} (c). Diffusion StyleGAN2 for CIFAR-10 and WGAN-GP
for MNIST are both poor generators since their generated samples can be
easily differentiated from real ones.  To further demonstrate the power
of the quality index, we show CIFAR-10 and MNIST generated examples with
soft scores close to 0 and 1 for visual comparison in Figs.
\ref{fig:examples_cifar} and \ref{fig:examples_mnist}, respectively.
Clearly, the soft score of a generated sample is correlated well with
its visual quality viewed by humans. 

\textbf{Post-processing of generative models.} This method can be
applied as a post-processing procedure for improving the quality of
generated samples by filtering out generated samples of bad quality.
Generated samples are sorted by their soft scores in ascending order and
good samples with smaller soft scores are kept as the post-processing
result. Fig. \ref{fig:discriminator} shows the accuracy and FID scores
with different percentages of kept generated samples. To keep the same
number of samples in accuracy and FID computation, more samples are used
for filtering out bad samples when the kept percentage is small. As the kept
percentage becomes smaller, the kept samples are of higher quality and
accuracy and the FID score improves. 

\textbf{Evaluation of generative models.} LGSQE provides evaluation
metrics for generative models by aggregating the quality indices of a
number of generative samples. For a generative model of good
performance, the real and generated samples will have a larger overlap
distribution region and thus more samples would have soft scores closer
to 0.5. Thus, an accuracy closer to 0.5 indicates a better generative
model. Table \ref{table:comp_eval} compares the performance metrics of
several methods in generative models evaluation, including (a) DCGAN,
(b) Diffusion-StyleGAN2 (c) StyleGAN2-ADA, (d) Styleformer, and (e)
StyleGAN-XL for CIFAR-10 (denoted by C) and (f) GAN, (g) WGAN, and (h)
WGAN-GP for MNIST (denoted by M). Acc. (accuracy), AUC (area under the curve), Pre. (precision) and Rec. (recall) are results of the proposed method. AP (average precision) is the AUC of
the precision-recall \cite{sajjadi2018assessing} curve, in which the
precision and recall are defined as quality and diversity metrics and a
higher AP means a better generative model. The performance ranking of
the proposed method is consistent with the most popular evaluation
metric, FID, and this further corroborates the effectiveness of our
method. Since AUC, Pre. and Rec. are close to AUC, these values aren't
computed for MNIST. 

\begin{table}[htb]
\caption{Performance comparison of several evaluation metrics.}
\label{table:comp_eval}
\centering
{\small
\resizebox*{\linewidth}{!}{
\begin{tabular}{|l|l||l|l|l|l|l|l|l|} \hline
DS. & Model & FID & IS & AP & Acc. & AUC & Pre. & Rec.\\ \hline
\multirow{5}*{C} & (a) & 47.7 & 6.58 & 0.86 & 0.95 & 0.99 & 0.95 & 0.95 \\
\cline{2-9} & (b) & 3.19 & - & 0.87 & 0.88 & 0.95 & 0.88 & 0.88  \\
\cline{2-9} & (c) & 2.92 & 9.83 & 0.87 & 0.84 & 0.92 & 0.84 & 0.84 \\
\cline{2-9} & (d) & 2.82 & 9.94 & 0.83 & 0.776 & 0.86 & 0.77 & 0.78 \\
\cline{2-9} & (e) & 1.85  & - & 0.79 & 0.62 & 0.68 & 0.62 & 0.65 \\ \hline
\multirow{3}*{M} & (f) & 26.56 & - & 0.72 & 0.99 & - & - & - \\
\cline{2-9} & (g) & 32.37 & - & 0.58 & 0.99 & - & - & - \\
\cline{2-9} & (h) & 26.12 & - & 0.65 & 0.76 & - & - & - \\ \hline
\end{tabular}
}}
\end{table}


\textbf{High model efficiency.} LGSQE has higher efficiency in terms of
its model size and inference time compared with other quality
evaluation metrics. The state-of-the-art metrics extract features from
Inception-v3, VGG-16, or ResNet-34 pretrained on ImageNet. Their model
sizes are 91.2MB, 138.4MB, and 83MB, respectively. However, it only takes 2-3 MB of memory for LGSQE to finish the whole evaluation process, including training and testing. It takes 122 minutes to finish FID computation on 10,000 pairs of generated and real images, while LGSQE only needs 2 to 3 minutes to complete the test under the
same computing setting.


\textbf{Setting of weakly supervised learning.} LGSQE can achieve
similar evaluation performance even with a smaller number of training
samples.  We show the test accuracy of LGSQE for CIFAR-10 as a function
of the number of training samples in Fig.  \ref{fig:diff_triannum} with
two generative models (i.e., Styleformer and Diffusion-StyleGAN). The
accuracy converges when 20\% real samples are used in training. In
contrast, FID and other evaluation metrics apply complex networks to
extract features. They demand more training data to get accurate
distribution statistics for the evaluation purpose. 

\section{Conclusion and Future Work} \label{sec:conclusion}

A lightweight evaluation metric, LGSQE, was presented to evaluate the
quality index of a generated sample. To the best of our knowledge, this
is the first quality metric developed for such a purpose.  It can
also be employed as an evaluation metric for generative models, maintaining
the same performance as other metrics, yet demanding fewer model
parameters, fewer training samples, and less inference time.  LGSQE can
be potentially used as a post-processing module for high quality image
generation. That is, it can reject low quality samples and, thus,
improve the overall image generation performance. This is an interesting
topic for future exploration. 

\bibliographystyle{IEEEtran}
\bibliography{001}

\end{document}